\documentclass[aps,prb,reprint,amsmath,amssymb,floatfix,superscriptaddress]{revtex4-2}

\usepackage{bm,graphicx,xcolor,multirow,hyperref}

\usepackage[ignoreunlbld,norefs,nocites]{refcheck}

\begin{document}

\title{Quantum Monte Carlo study of the phase diagram of the
  two-dimensional uniform electron liquid}

\author{Sam Azadi}

\email{sam.azadi@physics.ox.ac.uk}

\affiliation{Department of Physics, Clarendon Laboratory, University of Oxford, Parks Road, Oxford OX1 3PU, United Kingdom}

\author{N.\ D.\ Drummond}

\affiliation{Department of Physics, Lancaster University, Lancaster
  LA1 4YB, United Kingdom}

\author{Sam M.\ Vinko}

\affiliation{Department of Physics, Clarendon Laboratory, University of Oxford, Parks Road, Oxford OX1 3PU, United Kingdom}
\affiliation{Central Laser Facility, STFC Rutherford Appleton Laboratory, Didcot OX11 0QX, United Kingdom}

\date{\today}

\begin{abstract}
We present a study of spin-unpolarized and spin-polarized
two-dimensional uniform electron liquids using variational and
diffusion quantum Monte Carlo (VMC and DMC) methods with
Slater-Jastrow-backflow trial wave functions. Ground-state VMC and DMC
energies are obtained in the density range $1 \leq r_\text{s} \leq
40$.  Single-particle and many-body finite-size errors are corrected
using canonical-ensemble twist-averaged boundary conditions and
extrapolation of twist-averaged energies to the thermodynamic limit of
infinite system size. System-size-dependent errors in
Slater-Jastrow-backflow DMC energies caused by partially converged VMC
energy minimization calculations are discussed. We find that, for $1
\leq r_\text{s} \leq 5$, optimizing the backflow function at each
twist lowers the twist-averaged DMC energy at finite system
size. However, nonsystematic system-size-dependent effects remain in
the DMC energies, which can be partially removed by extrapolation from
multiple finite system sizes to infinite system size.  We attribute
these nonsystematic effects to the close competition between fluid and
defected crystal phases at different system sizes at low density. The
DMC energies in the thermodynamic limit are used to parameterize a
local spin density approximation correlation functional for
inhomogeneous electron systems. Our zero-temperature phase diagram
shows a single transition from a paramagnetic fluid to a hexagonal
Wigner crystal at $r_\text{s}=35(1)$, with no region of stability for
a ferromagnetic fluid.
\end{abstract}

\maketitle

The two-dimensional (2D) uniform electron liquid (UEL) is an important
model system in condensed matter physics and plasma physics
\cite{Giuliani05,Ceperley78,Tanatar,Gori-Giorgi,DePalo,Varsano,Attaccalite,Rapisarda96,Neilprl09,Neilprb09II}.
The 2D UEL model consists of a system of electrons moving in 2D in a
uniform, inert, neutralizing background, interacting via the Coulomb
potential. 2D UELs are realized in numerous semiconductor
heterostructures, e.g., a 2D UEL can be observed at the interface of
metal-oxide-semiconductor structures \cite{Ando},
metal-oxide-semiconductor field-effect transistors
\cite{Phillips98,Davies97}, quantum wells \cite{Datta95,Davies97}, and
MgZnO/ZnO heterostructures \cite{Nelson22}.

According to the 2D UEL phase diagram reported in
Ref.\ \onlinecite{Neilprl09}, the ground state is a non-spin-polarized
(paramagnetic) liquid from high density down to the point of Wigner
crystallization. However, in the absence of crystallization, lowering
the density would eventually make the paramagnetic fluid phase
unstable with respect to a spin-polarized (ferromagnetic) fluid. An
experimental study of nonequilibrium transport in low-density 2D
electron systems at zero external magnetic field suggests that a fully
spin polarized fluid is stable before crystallization \cite{Ghosh},
which was also reported by some previous theoretical studies
\cite{Tanatar,Rapisarda96}.  On the other hand, more recent
experimental work has not found a region of stability for the
ferromagnetic fluid \cite{Nelson22}. In this work we update the
real-space quantum Monte Carlo (QMC) calculations
reported in Ref.\ \onlinecite{Neilprl09} to include additional
long-range backflow terms in trial wave functions and we look at a
broader range of system sizes, resulting in a small modification to
the crystallization density.  Nevertheless, in agreement with the earlier
work, our present QMC calculations yield no density range for the
stability of fully or partially spin-polarized 2D UELs.

We have carried out QMC calculations for 2D UELs with spin
polarizations of $\zeta=0$, 0.5, and 1 within the density range $1
\leq r_\text{s} \leq 40$, where $r_\text{s}$ is the radius of the
circle that contains one electron on average in units of the Bohr
radius.  We have focused on reducing finite-size (FS) errors
\cite{Chiesa06,Drummond08,Holzmann16,Kent99} by using twist-averaging
(TA) and extrapolation techniques \cite{Ceperley77,Ceperley78,Lin01}.
In this work we discuss errors in DMC calculations near the
crystallization density introduced during wave function (WF)
optimization, whose effects may be exaggerated by FS extrapolation.

Slater-Jastrow-backflow (SJB) WFs \cite{Pablo06, Drummond04} with
plane wave orbitals $\exp(i\mathbf{G} \cdot \mathbf{r})$ were used in
our QMC calculations for fluid phases. Details of the SJB WF have been
explained in our recent studies \cite{Azadi23II,Azadi23,Azadi22}; the
principle difference from Ref.\ \onlinecite{Neilprl09} is the
inclusion of a long-range two-body backflow term ${\bm \pi}$ that is
expanded in a plane-wave basis.  This term lowers the variational
energy in finite cells, and is intended to make the treatment of
two-body correlations in different simulation cells more consistent,
to aid extrapolation to the thermodynamic limit. We used the
variational and diffusion quantum Monte Carlo (VMC and DMC) methods
\cite{Foulkes01} as implemented in the \textsc{casino} package
\cite{casino}. To impose fermionic antisymmetry in our DMC
calculations we have used the fixed-node approximation
\cite{Anderson}, where the nodal surface is forced to be the same as
that of the trial WF\@. The variational parameters in the Jastrow
factor and backflow functions were optimized within VMC by minimizing
either the energy variance \cite{Umrigar88,Neil05} or the mean
absolute deviation of the local energies from the median
\cite{Bressanini_2002}, then minimizing the energy expectation value
\cite{Umrigar88,Umrigar}. We used a target population of 2560 walkers
in our DMC calculations. At $r_\text{s}=1$, 2, 5, 10, 15, 20, 25, 30,
35, and 40 we used DMC time steps of $\tau=0.005$, 0.01, 0.05, 1, 1.5,
2, 3, 4, 5, and 6 Ha$^{-1}$, respectively.  Random errors due to TA
are larger than the time-step bias in each case.  Our DMC energies
obtained at various system sizes using a hexagonal simulation cell are
reported in the Supplemental Material \cite{Suppl}.

Except where otherwise stated, our Jastrow factors and backflow
functions were optimized at zero twist (i.e., pure periodic boundary
conditions).  For paramagnetic 2D UELs with densities $r_\text{s}=1$,
2, and 5, and system size $N=146$, we tested separately optimizing the
Jastrow factor and backflow function at each twist. The results are
listed in Table~\ref{tab:opttwist}.  We have used 30 random twists in
our analysis.  Our analysis shows that separately optimizing the
Jastrow factor and backflow function at each twist reduces the TA DMC
energy by $-0.225(9) \times 10^{-3}/r_\text{s}$ Ha/el.\ for
paramagnetic 2D UELs at $N=146$. These results demonstrate the
existence of significant, quasirandom FS errors in TA energies in
which the same backflow function is used for all twists.

\begin{table}
\centering
\caption{TA DMC energies of paramagnetic 146-electron 2D UELs.  The
  DMC calculations were performed using either the same Jastrow factor
  and backflow function, optimized at zero twist, for all twists
  (DMC$_0$) or SJB wave functions separately optimized at each twist
  (DMC$_1$). \label{tab:opttwist}}
 \begin{tabular}{lccc}
 \hline\hline

 \multirow{2}{*}{$r_\text{s}$} & \multicolumn{2}{c}{TA DMC energy (Ha/el.)}
 & \multirow{2}{*}{Difference (Ha/el.)} \\

 & DMC$_0$ & DMC$_1$ & \\

 \hline
 1  & $-0.210731(7)$ & $-0.210957(5)$ & $-0.225(8)\times 10^{-3}$\\
 2  & $-0.258378(9)$ & $-0.258501(6)$ & $-0.12(1)\times 10^{-3}$ \\
 5  & $-0.149595(4)$ & $-0.149637(2)$ & $-0.042(5)\times 10^{-3}$  \\
 \hline\hline
  
 \end{tabular}
\end{table}

WF optimization is challenging at large system sizes ($N \gtrsim 150$)
and at densities near the solid-liquid phase transition ($r_\text{s}
\gtrsim 20$) \cite{Suppl}. Our final WFs are obtained by energy
minimization, but usually this requires a reasonable first
approximation to the WF, obtained using a different method.  Although
unreweighted variance minimization exhibits superior numerical
stability to energy minimization at high and intermediate densities,
we found numerical instabilities in variance minimization for systems
with large $N$ and low density. Numerical instability in variance
minimization can occur when the nodes of the trial WF are altered
during minimization.  Near the transition density we find that
numerical instabilities become more severe at large system size due to
the increased likelihood of producing a wave function for the
``wrong'' phase (e.g., a defected crystal wave function instead of a
fluid wave function) and the growth of complexity of the energy
landscape in parameter space.  Minimizing a more robust measure of the
spread of the local energies, such as the mean absolute deviation
(MAD) from the median local energy \cite{Bressanini_2002}, provides a
much lower variational energy (Table~\ref{tab:optparam}), because the
MAD is less sensitive to divergent local energies \cite{casino}.

\begin{figure}[htbp!]
    \centering
    \includegraphics[clip,scale=0.65]{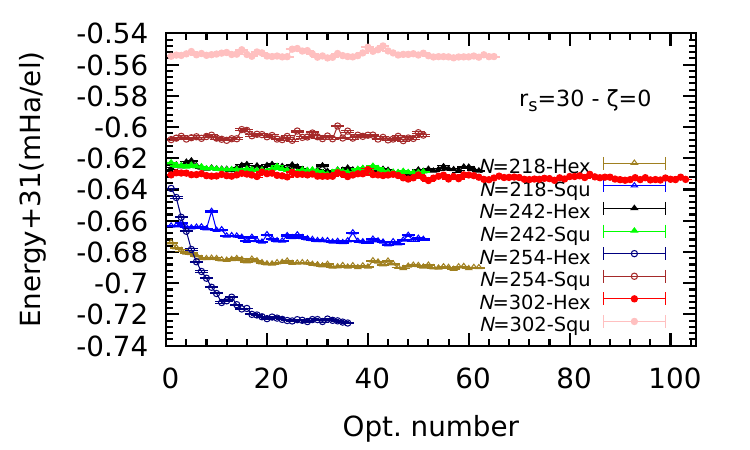}
    \includegraphics[clip,scale=0.65]{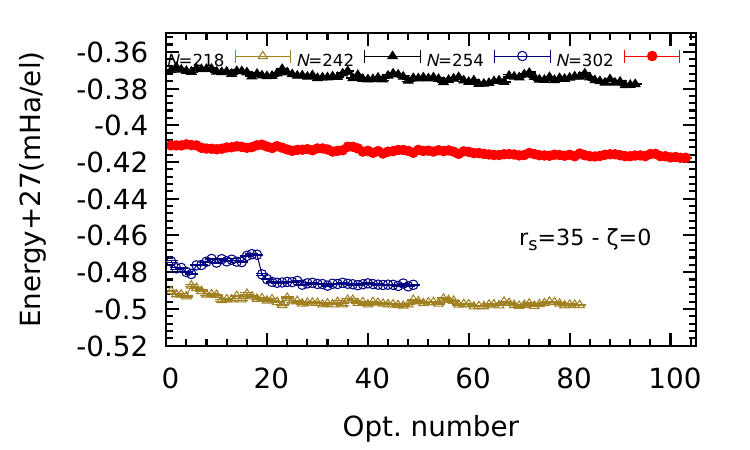}    
    \caption{SJB VMC energy against energy minimization cycle for
      paramagnetic 2D UELs with density parameters $r_\text{s}=30$
      (top panel) and 35 (bottom panel). The systems studied had
      $N=218$, 242, 254, and 302 electrons. The initial WF was
      optimized by minimizing the mean absolute deviation from the
      median local energy.  The real WF was optimized at zero
      twist. The energies of the 2D UELs at $r_\text{s}=30$ were
      obtained using hexagonal and square simulation cells.}
    \label{fig:eminopt}
\end{figure}

\begin{figure}
     \centering
    \includegraphics[clip,scale=0.68]{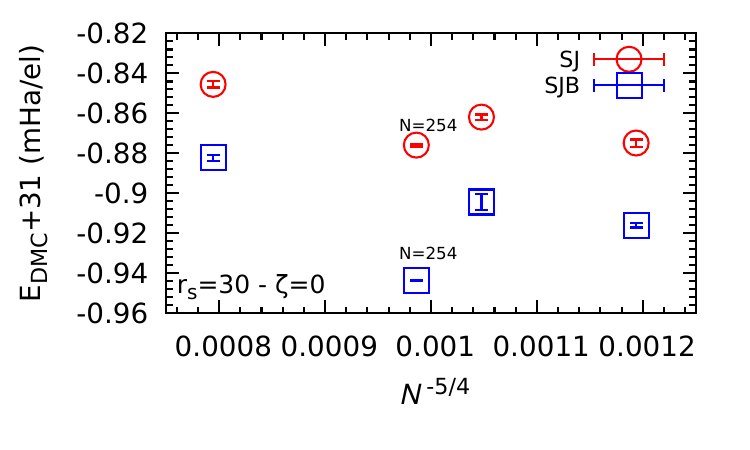}
    \caption{TA DMC energies for paramagnetic ($\zeta=0$) 2D UELs with
      density parameter $r_\text{s}=30$ obtained using SJ and SJB wave
      functions. The differences between the SJB and SJ energies for
      $N=218$, 242, 254, and 302 are $-0.041(2)$, $-0.042(4)$,
      $-0.068(1)$, and $-0.037(2)$ mHa/el., respectively.}
    \label{fig:DMCSJrs30}
\end{figure}

\begin{figure}[htbp!]
    \centering
    \includegraphics[clip,scale=0.65]{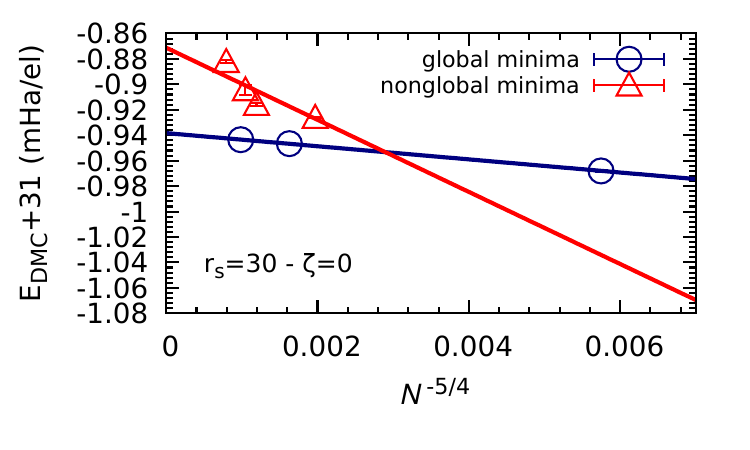}
    \caption{Extrapolation of TA DMC fluid energies to the
      thermodynamic limit of infinite system size for paramagnetic
      ($\zeta=0$) 2D UELs with density parameter $r_\text{s}=30$. The
      red data points are assumed to be trapped in nonglobal minima
      during optimization and do not correspond to the fluid ground
      state.}
    \label{fig:DMCminima}
\end{figure}

We found that more than a dozen cycles of VMC configuration-generation
and energy minimization are often required to converge adequately in
the regime of large $N$ and large $r_\text{s}$, as shown in
Fig.\ \ref{fig:eminopt}.  The magnitude of the effect is significant
on the scale of the FS error, even if the subsequent use of DMC
weakens the dependence on the backflow function.  A related
observation is that the TA VMC and DMC energies per particle are often
higher than suggested by extrapolation from smaller system sizes (see
Figs.\ \ref{fig:DMCSJrs30} and \ref{fig:DMCminima}).  In finite UELs
of a given density and spin polarization, either Wigner crystal-like
or Fermi fluid-like states may be energetically favored at different
system sizes $N$, different simulation cell shapes, and different
twists.  Fluid energies fluctuate as a function of $N$ due to
shell-filling effects in reciprocal space, while crystal energies
fluctuate as a function of $N$ due to spatial commensurability effects
in real space.  Furthermore, there may be many near-degenerate
low-energy states, e.g., due to different defected Wigner crystals in
a particular cell.  Even if a fluid trial wave function is used, it
may be energetically advantageous for the system to approximate a
floating Wigner crystal when the wave function is optimized.
Conversely, at particular supercell sizes, it may be advantageous for
a Wigner crystal wave function to try to approximate a fluid excited
state.  For certain simulation cells it may be especially difficult to
find the lowest-energy fluid or crystal wave function.  We can try to
tip the balance in favor of our selected phase by using ``magic''
numbers of electrons (such that the occupied plane-wave orbitals form
a closed-shell configuration in reciprocal space) for fluid phases,
and by using square numbers of electrons for crystals, such that
defect-free lattices are commensurate with the simulation cell.
However, this is not guaranteed to prevent wave-function optimization
giving an approximation to an excited state of the wrong phase.
Although a Jastrow factor $\exp(J)$ cannot truly introduce nodes, if
$J$ becomes large and negative then it may introduce ``quasinodes''
that affect the DMC energy over any reasonable imaginary time scale.

Figure \ref{fig:eminopt} shows the energy minimization process for the
paramagnetic liquid phase with system sizes $N=218$, 242, 254, and 302
and densities of $r_\text{s}=30$ and 35. The SJB WFs used for energy
minimization were initially optimized by MAD minimization. The problem
of slow convergence with optimization cycle can be observed in
Fig.\ \ref{fig:eminopt} and in the Supplemental Material \cite{Suppl}.
These figures also show occasional jumps suggestive of wave-function
parameters moving from one local minimum-energy configuration to
another.

Residual canonical-ensemble TA errors in the Hartree--Fock kinetic and
exchange energies around the crystallization density are orders of
magnitude smaller than the fluctuations in the DMC energy shown in
Fig.\ \ref{fig:DMCminima}; hence residual momentum quantization
effects in the TA energy cannot explain the nonsystematic system-size
dependence.  Other sources of nonsystematic FS error, such as
Ruderman-Kittel oscillations being forced to be commensurate with the
simulation cell, may be present in Fig.\ \ref{fig:DMCSJrs30}.  Such
nonsystematic FS errors are at least partially averaged out by
extrapolation to the thermodynamic limit.

\begin{figure}[htbp!]
    \centering
    \includegraphics[clip,scale=0.34]{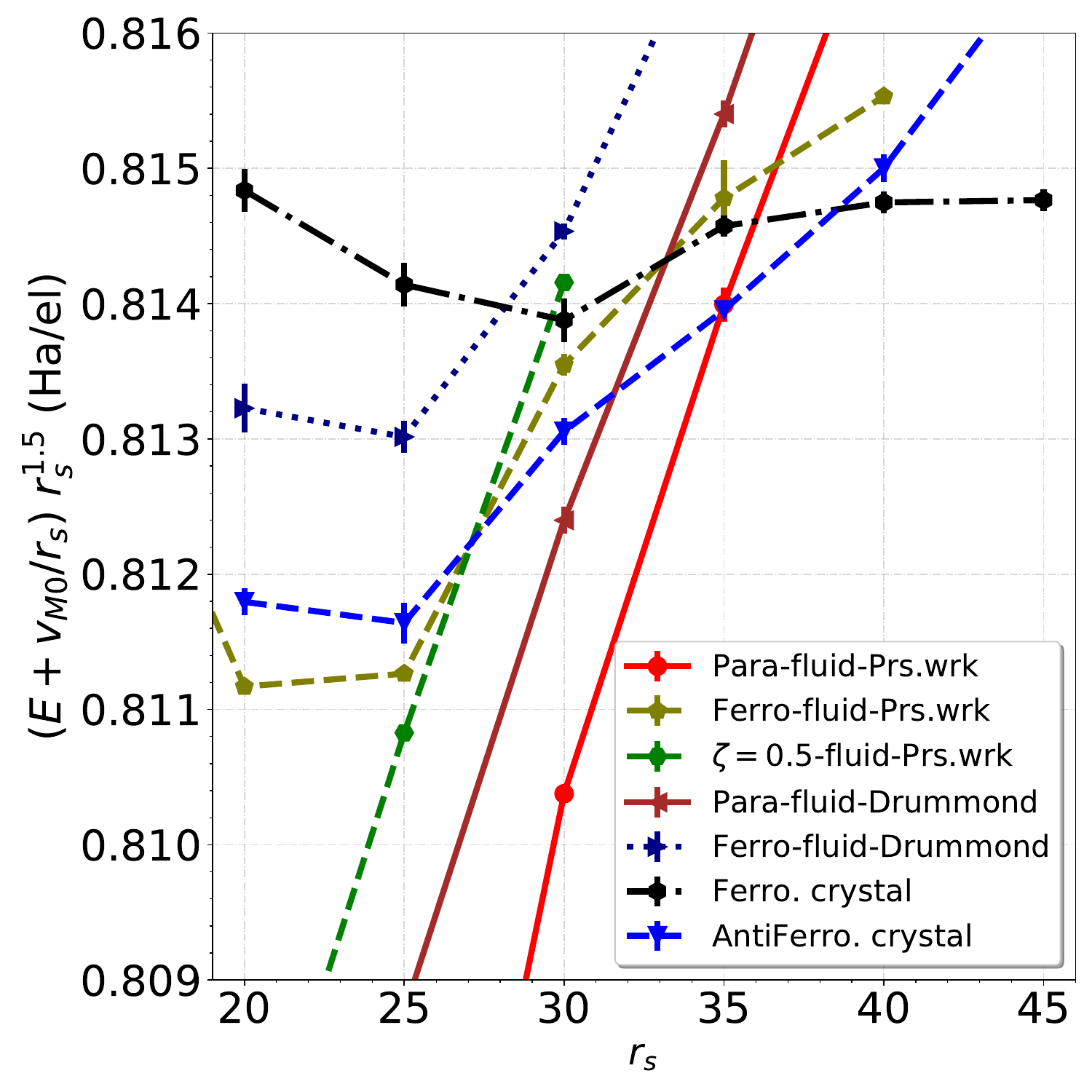}
    \caption{DMC energy extrapolated to infinite system size as a
      function of $r_\text{s}$ for 2D UELs with spin polarizations
      $\zeta=0$, 0.5, and 1. Our results (``Prs-wrk'') are compared
      with those of Drummond and Needs \cite{Neilprl09}, which include
      paramagnetic ($\zeta=0$) and ferromagnetic fluids ($\zeta=1$),
      and ferromagnetic and antiferromagnetic hexagonal Wigner
      crystals. The Madelung energy of a hexagonal lattice,
      $-v_\text{M0}/r_\text{s}$, where
      $v_\text{M0}=-0.50751467391482663$ is the Madelung constant at
      $r_\text{s}=1$, has been subtracted from all the DMC
      energies.}
    \label{fig:phasediagram}
\end{figure}

We investigated this issue further. First, we investigated how the
initial SJB WF optimization depends on the number of configurations
and the twist. The results are summarized in Table~\ref{tab:optparam}.
They indicate that the optimized energy for a large system at low
density does not depend strongly on the number of configurations or
the twist wavevector, but does depend on the optimization method.
Because the local energy diverges at nodes, the unweighted variance
and MAD landscapes depend strongly on the sampled configurations.
Nevertheless, it is clear that MAD minimization provides much better
results than variance minimization and hence better starting points
for energy minimization. Second, we compared the FS behavior with
Slater-Jastrow (SJ) and SJB WFs. The fixed-node SJ-DMC energy is
independent of the Jastrow factor.  Figure \ref{fig:DMCSJrs30} shows
that the TA SJ-DMC energies behave in a nonsystematic manner as a
function of system size, and that this behavior is further exaggerated
by the inclusion of backflow.  This demonstrates (i) that the
nonsystematic behavior is not simply a result of difficulties
optimizing wave functions and (ii) that applying analytic FS
corrections \cite{Chiesa06,Drummond08,Holzmann16} to results obtained
at a single fixed cell size is unreliable because nonsystematic FS
effects cannot be removed by this approach.  Third, we examined the
static structure (SF) factor and momentum-density (MD) of a
paramagnetic 2D UEL with $N=254$ at $r_\text{s}=30$ obtained by VMC at
a single twist.  The results are presented in the Supplemental
Material \cite{Suppl}.  We have not found significant
WF-optimization-dependent or system-size dependent anomalies in the SF
and MD (e.g., the MD retains its discontinuity at the Fermi
wavevector); unfortunately one cannot straightforwardly check for the
wrong phase emerging, as the SJB wave function does not truly permit a
change of phase. Fourth, we compared the energies of 2D UELs at
$r_\text{s}=30$ obtained using hexagonal and square simulation cells
with system sizes $62 \leq N \leq 302$ \cite{Suppl}.
The Wigner crystal energies used in our phase diagram
(Fig.\ \ref{fig:phasediagram}) were calculated using hexagonal
simulation cells.  Since the Madelung energy of a square lattice is
higher, using a square lattice would have made crystal-like wave
functions less competitive.
Fifth, we performed SJ-VMC, SJ-DMC, and DMC with only the Slater trial
wave function (S-DMC) calculations for paramagnetic Fermi fluids at
$r_\text{s}=30$ in square cells with $N=90$ electrons.  We chose three
different initial wave functions for the optimization (Starting points
1--3) and optimized the Jastrow factor by energy minimization at
either $\Gamma$ (${\bf k}_\text{s}={\bf 0}$) or the Baldereschi point
[${\bf k}_\text{s} = (1/4) {\bf b}_1 + (1/4) {\bf b}_2$, where ${\bf
    b}_1$ and ${\bf b}_2$ are the supercell reciprocal lattice
  vectors].  The DMC calculations were performed at time steps of 2
and 8 Ha$^{-1}$, with populations in inverse proportion to the time
step, and the energies were extrapolated linearly to zero time step
and infinite population.  Hartree-Fock and VMC energies are shown in
Table \ref{tab:sq_cell_fluid_E_VMC}, while DMC energies with and
without a Jastrow factor are shown in Table
\ref{tab:sq_cell_fluid_E_DMC}.

\begin{table*}[htbp!]
 \centering
 \caption{\label{tab:optparam}
   Optimized energy of paramagnetic 2D UELs at $r_\text{s}=30$ with
   $N=302$ electrons. Only polynomial and plane-wave two-body Jastrow
   terms were used. The energies were calculated using different
   numbers of configurations $N_\text{conf}$. Two different
   optimization methods, unreweighted variance minimization
   (``varmin'') and MAD minimization (``madmin''), were used. Three
   twists, with fractional coordinates ${\bf k}_0=(0,0)$, ${\bf
     k}_1=(1/3,1/3)$, and ${\bf k}_2=(1/4,1/2)$, were used. All the
   energies are in Ha/el. The energies which are lower than $-0.031$
   Ha/el.\ are highlighted in bold.}
 \begin{tabular}{lcccccc}
  \hline\hline
  \multirow{2}{*}{$N_\text{conf}$} & \multicolumn{2}{c}{Twist ${\bf k}_0$} & \multicolumn{2}{c}{Twist ${\bf k}_1$} & \multicolumn{2}{c}{Twist ${\bf k}_2$} \\
  & ``varmin'' & ``madmin'' & ``varmin'' & ``madmin'' & ``varmin'' & ``madmin'' \\
  \hline
  $1920$ & $-0.027955(7)$& $-0.019491(8)$& $-0.025725(2)$& ${\bf -0.0315169(4)}$ &$-0.025009(2)$ & ${\bf -0.0314244(5)}$ \\
  $3840$ & $-0.019497(8)$& $-0.019508(8)$& $-0.028892(8)$& $-0.019487(9)$ & $-0.0307900(8)$ & ${\bf -0.0315429(4)}$ \\ 
  $7680$ &$-0.0307833(9)$ & ${\bf -0.0314710(4)}$&$-0.01950(1)$ &$-0.019496(9)$ & $-0.029414(3)$& ${\bf -0.0315166(4)}$ \\
  $9600$ & $-0.025907(2)$& ${\bf -0.0315087(4)}$& $-0.019502(9)$& $-0.019496(8)$ & $-0.028948(2)$& $-0.01948(1)$ \\
  $11520$&$-0.019503(8)$ & ${\bf -0.0315034(4)}$& $-0.030772(3)$& ${\bf -0.0315255(4)}$& $-0.0308548(6)$& $-0.01949(1)$\\
  $15360$& $-0.019492(8)$&$-0.019523(9)$ & ${\bf -0.0313126(5)}$& $-0.01948(2)$ &$-0.030773(1)$ & $-0.019507(9)$ \\
  $19200$&$-0.027983(2)$ &$-0.019519(8)$ &$-0.030738(2)$ &$-0.019494(9)$ &$-0.01947(1)$ & ${\bf -0.0315176(5)}$ \\
  \hline\hline
 \end{tabular}

\end{table*}

\begingroup
\squeezetable
\begin{table}
\centering
\caption{Hartree-Fock (HF) and SJ-VMC energies for 90-electron Fermi
  fluids in a square simulation cell, with the wave function being
  optimized from three different random starting points at the
  $\Gamma$ point and at the Baldereschi point of the square simulation
  cell. \label{tab:sq_cell_fluid_E_VMC}}
\begin{tabular}{lccccc}
\hline \hline

\multirow{2}{*}{${\bf k}_\text{s}$} & \multirow{2}{*}{Start} & \multicolumn{4}{c}{Energy (Ha/el.)} \\

& & HF & TA HF & SJ-VMC & TA SJ-VMC \\

\hline

$\Gamma$ & 1 & $-0.0196518$ & $-0.0196034$ & $-0.0317373(2)$ &
$-0.0317344(2)$ \\

$\Gamma$ & 2 & $-0.0196518$ & $-0.0196034$ & $-0.0251398(8)$ &
$-0.025177(2)$ \\

$\Gamma$ & 3 & $-0.0196518$ & $-0.0196034$ & $-0.0271506(7)$ &
$-0.02716(1)$ \\

Bald. & 1 & $-0.0194515$ & $-0.0196034$ & $-0.0317356(2)$ &
$-0.0317350(2)$ \\

Bald. & 2 & $-0.0194515$ & $-0.0196034$ & $-0.02495(1)$ &
$-0.02500(1)$ \\

Bald. & 3 & $-0.0194515$ & $-0.0196034$ & $-0.0277466(6)$ &
$-0.027726(7)$ \\

\hline \hline
\end{tabular}
\end{table}
\endgroup

\begingroup
\squeezetable
\begin{table}
\centering
\caption{As Table \ref{tab:sq_cell_fluid_E_VMC}, but using S-DMC and
  SJ-DMC calculations. \label{tab:sq_cell_fluid_E_DMC}}
\begin{tabular}{lccccc}
\hline \hline

\multirow{2}{*}{${\bf k}_\text{s}$} & \multirow{2}{*}{Start} &
\multicolumn{4}{c}{Energy (Ha/el.)} \\

& & S-DMC & TA S-DMC & SJ-DMC & TA SJ-DMC \\

\hline

$\Gamma$ & 1 & $-0.03151(2)$ & $-0.03151(2)$ & $-0.031919(1)$ &
$-0.031903(1)$ \\

$\Gamma$ & 2 & $-0.03151(2)$ & $-0.03151(2)$ & $-0.02676(7)$ &
$-0.02683(4)$ \\

$\Gamma$ & 3 & $-0.03151(2)$ & $-0.03151(2)$ & $-0.02848(2)$ &
$-0.02849(2)$ \\

Bald. & 1 & $-0.03151(2)$ & $-0.03159(2)$ & $-0.031899(2)$ &
$-0.031898(1)$ \\

Bald. & 2 & $-0.03151(2)$ & $-0.03159(2)$ & $-0.02670(2)$ & \\

Bald. & 3 & $-0.03151(2)$ & $-0.03159(2)$ & $-0.02901(2)$ &
$-0.02901(2)$ \\

\hline \hline
\end{tabular}
\end{table}
\endgroup

As expected, the choice of twist for optimization has only a small
effect on the fluid energy at this low density.  Indeed, twist
averaging only has a small effect on the fluid energy.  Starting
points 2 and 3 produce wave functions that give a significantly lower
VMC energy than the Hartree-Fock energy, i.e., they introduce
correlations that lower the energy.  But the energy obtained is not as
low as the energy obtained in starting point 1.  Hence there are
issues with local minima.

The S-DMC energy obtained with the HF wave function is lower than the
SJ-DMC energy obtained using the wave functions from starting points 2
and 3, despite the fact that the HF energy is higher than the SJ-VMC
energies.  Hence these Jastrow factor introduce features
(``quasinodes'') that are difficult for DMC to remove (i.e., would
require much smaller time steps).  Since the HF wave function clearly
corresponds to a fluid phase, it is reasonable to conclude that the SJ
wave functions from starting points 2 and 3 do not correspond to a
fluid phase.  The S-DMC energy is in fairly good agreement with the
SJ-DMC energy obtained with the good Jastrow factor obtained in
starting point 1.  The remaining small difference is very likely
because of the need to extrapolate the S-DMC energy to zero time step
a bit more carefully.  Hence the Jastrow factor obtained in starting
point 1 is very likely to describe a fluid phase.

Overall the picture is consistent with the idea that optimization of
the wave function can get trapped in local minima that are not
fluid-like, and DMC is unable to repair the situation when this
happens.  Some system sizes are more prone to problems with wave
function optimization than others, due to the relative availability of
crystal-like local minima.  A disturbing feature is that one can
optimize a wave function and lower the VMC energy, and yet the DMC
energy is worse than not having a Jastrow factor at all.  Having a
more complex trial wave function is not a guarantee that the
calculation is improved: it can make it easier to fall into unwanted
local minima.  All these conclusions apply more strongly in the case
of SJB wave functions, which directly affect the DMC energy, even in
the limit of zero time step.

We calculated the TA DMC energy for all the system sizes shown in
Fig.\ \ref{fig:eminopt}.  The energy at the thermodynamic limit is
obtained by a linear extrapolation of the TA DMC energy as a function
of $N^{-5/4}$ \cite{Neilprb09II}. Figure \ref{fig:DMCminima} shows the
extrapolation of TA DMC energies of two sets of data points labeled as
``nonglobal minima'' and ``global minima.''  All the TA DMC energies
at the infinite system size limit presented in the rest of this paper
are obtained using the extrapolation of TA DMC of systems that we
believe correspond to the fluid ground state.  

Table \ref{tab:allenrgies} lists our TA DMC energies at the
thermodynamic limit for 2D UELs with spin polarizations $\zeta=0$,
0.5, and 1, and density parameters $1 \leq r_\text{s} \leq 40$.  We
compare our results with the previous works of Drummond and Needs
\cite{Neilprl09,Neilprb09II}.  The energies extrapolated to infinite
system size are slightly lower than those obtained by Drummond and
Needs. There are two major differences between our work and the
previous ones.  First, we used the two-body plane-wave backflow ${\bm
  \pi}$ term \cite{Azadi23, Azadi23II}, giving additional variational
freedom in simulation cells of finite size. Second, we used larger
system sizes and hexagonal simulation cells.  These differences do not
matter in principle, but in practice they affect the extrapolation to
infinite system size.

\begin{table*}[htbp!]
    \centering
    \caption{DMC energy in Ha/el.\ in the infinite system-size limit
      for 2D UELs with spin polarizations $\zeta=0$, 0.5, and 1 and
      density parameters $1 \leq r_\text{s} \leq 40$. The results of
      Drummond and Needs are taken from Refs.\ \onlinecite{Neilprl09}
      and \onlinecite{Neilprb09II}. \label{tab:allenrgies}}
    \begin{tabular}{lccccc}
    \hline\hline
    \multirow{2}{*}{$r_\text{s}$} & \multicolumn{3}{c}{Present work} & \multicolumn{2}{c}{Drummond and Needs} \\
 & $\zeta=0$ &  $\zeta=0.5$  & $\zeta=1$ & $\zeta=0$ &  $\zeta=1$ \\
    \hline
    1   & $-0.21017(6)$  & \ldots        &  $0.12626(4)$ & $-0.2104(6)$    &\ldots       \\
    2   & $-0.25846(7)$  & \ldots        & $-0.19453(1)$ & \ldots        & \ldots      \\
    5   & $-0.14998(2)$  & \ldots        & $-0.143714(2)$& $-0.14963(3)$   & \ldots      \\  
    10  & $-0.085568(5)$  & \ldots       & $-0.084585(6)$ & $-0.085399(6)$ & \ldots      \\   
    15  & $-0.06138(2)$   & \ldots       & $-0.059731(1)$ & \ldots       & \ldots      \\
    20  & $-0.046388(1)$  & $-0.0462813(1)$& $-0.046236(1)$ & $-0.046305(4)$ & $-0.046213(3)$  \\
    25  & $-0.037807(3)$  & $-0.0377575(1)$& $-0.037754(1)$ & $-0.037774(2)$ & $-0.037740(2)$  \\
    30  & $-0.0319383(2)$ & $-0.0319153(1)$& $-0.031919(2)$ & $-0.031926(1)$ & $-0.031913(1)$  \\
    35  & $-0.0276718(4)$ & \ldots       & $-0.027668(7)$ & $-0.027665(1)$ & $-0.027657(1)$  \\
    40  & $-0.0244226(3)$ & \ldots       & $-0.0244289(1)$& $-0.024416(1)$ & $-0.024416(1)$  \\
    \hline\hline
    \end{tabular}
\end{table*}

We used our DMC energies extrapolated to infinite system size to
calculate the phase diagram of the 2D UEL
(Fig.\ \ref{fig:phasediagram}).  The ferromagnetic and
antiferromagnetic crystal energies used in our phase diagram
calculation are taken from Ref.\ \onlinecite{Neilprl09}.  We find that
the paramagnetic Fermi fluid transitions to a hexagonal Wigner crystal
at $r_\text{s}=35(1)$.  The previous work by Drummond and Needs
\cite{Neilprl09} predicted a value of $r_\text{s}=31(1)$ for this
phase transition.  Similar to the previous phase diagram predicted by
Drummond and Needs \cite{Neilprl09}, we find no region of stability
for the polarized fluid phase with spin polarizations of $\zeta=0.5$
or 1.

The correlation energy $E_\text{c}$ is approximately given by the
difference between the TA DMC energy at the thermodynamic limit and
the Hartree-Fock energy. Our correlation energy results are shown in
Table \ref{tab:CorrEner}.
\begin{table}[htbp!]
    \centering
    \caption{SJB-DMC correlation energy in the limit of infinite
      system size for 2D UELs with spin polarizations $\zeta=0$ and
      1. \label{tab:CorrEner}}
    \begin{tabular}{lcc}
    \hline\hline
    \multirow{2}{*}{$r_\text{s}$} & \multicolumn{2}{c}{Correlation energy (Ha/el.)} \\
     & $\zeta=0$  & $\zeta=1$ \\
    \hline
     1 & $-0.10996(6)$  & $-0.02491(4)$ \\
     2 & $-0.08335(7)$  & $-0.02012(1)$ \\
     5 & $-0.04993(2)$  & $-0.013948(2)$ \\
     10& $-0.03054(5)$  & $-0.009702(6)$ \\
     15& $-0.02359(2)$  & $-0.007587(1)$ \\
     20& $-0.017627(1)$ & $-0.006295(1)$ \\
     25& $-0.014598(3)$ & $-0.005401(1)$ \\
     30& $-0.012487(2)$ & $-0.004736(2)$ \\
     35& $-0.0109311(4)$& $-0.004232(7)$ \\
     40& $-0.0097298(3)$& $-0.0038332(1)$ \\
     \hline\hline
    \end{tabular}
\end{table}
We fit our TA DMC correlation energies to the Pad\'{e} function \cite{Tanatar}
\begin{equation}
    E_\text{c}(r_\text{s}, \zeta)=a_\zeta\times\frac{1+b_\zeta r_\text{s}^{1/2}}{1+b_\zeta r_\text{s}^{1/2}+c_\zeta r_\text{s} + d_\zeta r_\text{s}^{3/2}}, \label{eq:corr_en_fit}
\end{equation}
which has the correct asymptotic behavior at high and low
densities \cite{Tanatar}, i.e., $E_\text{c}(r_\text{s}) = a -ac r_\text{s} +
O(r_\text{s}^{3/2})$ at small $r_\text{s}$, while
$E_\text{c}(r_\text{s}\rightarrow \infty) =
(ab/d)r_\text{s}^{-1}+a(1/d-bc/d^2)r_\text{s}^{-3/2}+O(r_\text{s}^{-2})$
at large $r_\text{s}$.  The fitting parameters $a$, $b$, $c$, and $d$
are listed in Table \ref{tab:CorrParam}.

\begin{table}[htbp!]
    \centering
    \caption{Fitting parameters for the SJB-DMC correlation energy
      [Eq.\ (\ref{eq:corr_en_fit})] of paramagnetic ($\zeta=0$) and
      ferromagnetic ($\zeta=1$) 2D UELs\@. \label{tab:CorrParam}}
    \begin{tabular}{lcc}
    \hline\hline
    Parameter  & $\zeta=0$ & $\zeta=1$ \\ 
    \hline
     $a$ (Ha/el.) & $-0.1842$ & $-0.0426$ \\
     $b$  & $3.2320$  & $28.8732$ \\
     $c$  & $1.6027$  & $16.5990$ \\
     $d$  & $1.2541$  & $4.6858$ \\
    \hline\hline
    \end{tabular}
\end{table}

In summary, we have used SJB WFs to perform VMC and DMC calculations
for 2D UELs with spin polarizations $\zeta=0$, 0.5, and 1 within the
density range $1\leq r_\text{s} \leq 40$. We have corrected
single-particle and many-body FS errors by canonical-ensemble TA and
extrapolation of the TA energies to infinite system size. We find that
separately optimizing the Jastrow factor and backflow function at each
twist improves the TA DMC energy for the density range $1\leq
r_\text{s} \leq 5$ and possibly beyond.  Optimization of the trial
wave function is challenging near the transition density because the
complexity of the energy landscape makes it difficult to find the
ground state corresponding to the Fermi fluid.  We predict that the
paramagnetic 2D UEL transforms to a Wigner crystal at
$r_\text{s}=35(1)$. We have not found any region of stability for 2D
UELs with spin polarizations $\zeta=0.5$ or 1.

\bibliography{main}

\begin{thebibliography}{35}
\bibitem [{Sup()}]{Suppl}%
  \BibitemOpen
  \href@noop {} {\bibinfo {title} {Raw quantum {M}onte {C}arlo data for all the
  systems studied are presented in the {S}upplemental {M}aterial}}\BibitemShut
  {NoStop}%
\end{thebibliography}

\end{document}


\title{Quantum Monte Carlo study of the phase diagram of the
  two-dimensional uniform electron liquid (Supplemental Material)}

\author{Sam Azadi}

\email{sam.azadi@physics.ox.ac.uk}
\affiliation{Department of Physics, Clarendon Laboratory, University of Oxford, 
Parks Road, Oxford OX1 3PU, United Kingdom}

\author{N.\ D.\ Drummond}

\affiliation{Department of Physics, Lancaster University, Lancaster LA1 4YB, United Kingdom}

\author{Sam M.\ Vinko}

\affiliation{Department of Physics, Clarendon Laboratory, University
  of Oxford, Parks Road, Oxford OX1 3PU, United Kingdom}

\date{\today}

\maketitle

\begin{figure}[htbp!]
    \centering
     \includegraphics[scale=0.62,angle=-90]{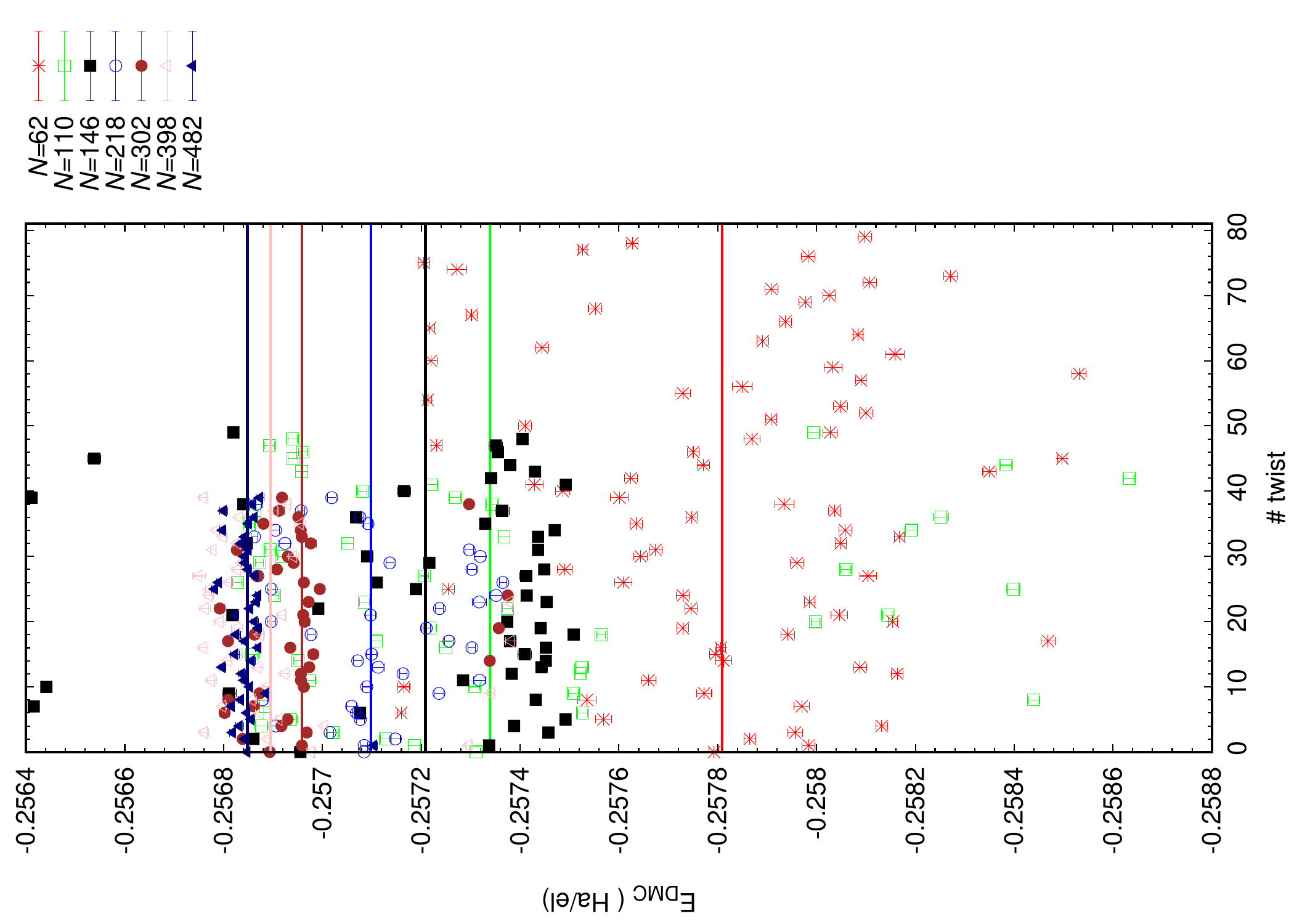}
 \caption{\label{tavenergy} Diffusion Monte Carlo energy
   $E_\text{DMC}$ against randomly sampled twist for two-dimensional
   uniform electron liquids with density parameter $r_\text{s}=2$ and
   polarization $\zeta=0$ with a Slater-Jastrow wave
   function. Energies are obtained for different system sizes $N$ and different
   number of twists. 80
   random twists are used for $N=62$ and smaller number of random twists are used for
   the other system sizes.}
\end{figure}

\begin{figure}[htbp!]
    \centering
            \begin{tabular}{cc}

    \includegraphics[scale=0.69]{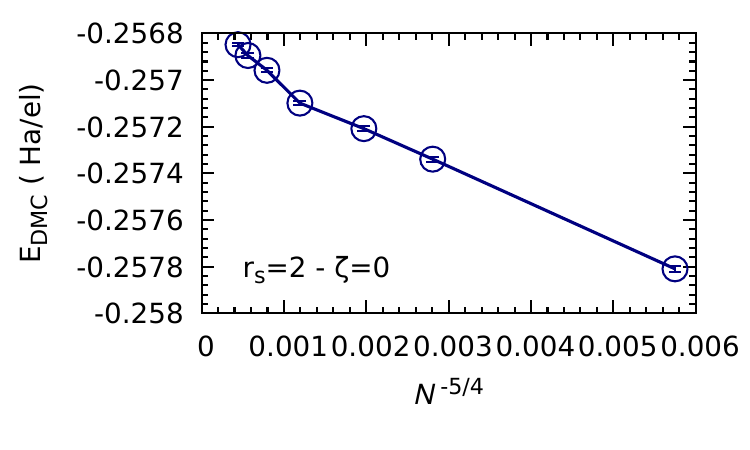} &
    \includegraphics[scale=0.69]{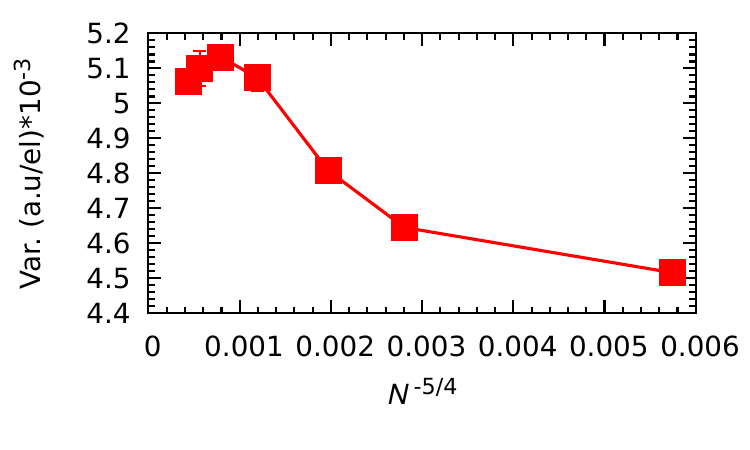} 
                \end{tabular}

    \caption{(Left panel) Twist-averaged diffusion Monte Carlo energy
      $E_\text{DMC}$ with a Slater-Jastrow wave function as a function
      of system size for paramagnetic two-dimensional uniform electron
      liquids with $r_\text{s}=2$. (Right panel) Variational quantum
      Monte Carlo energy variance at zero twist as a function of
      system size.}
    \label{fig:DMCTAVrs2}
\end{figure}

\begin{figure}[htbp!]
        \centering
        \begin{tabular}{cc}
    \includegraphics[scale=0.69]{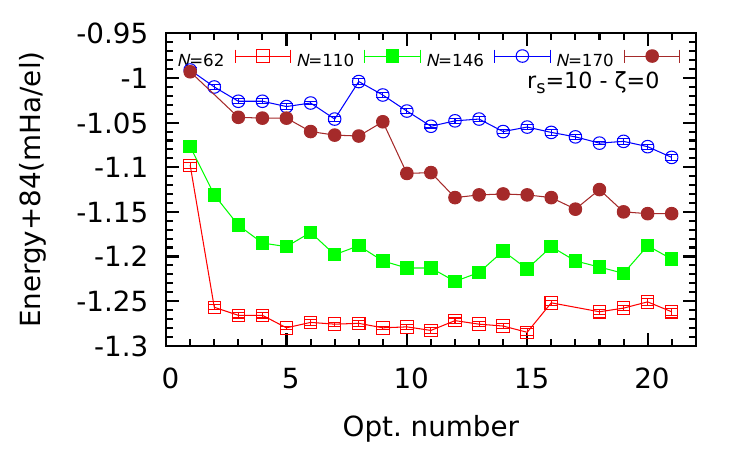} &
    \includegraphics[scale=0.69]{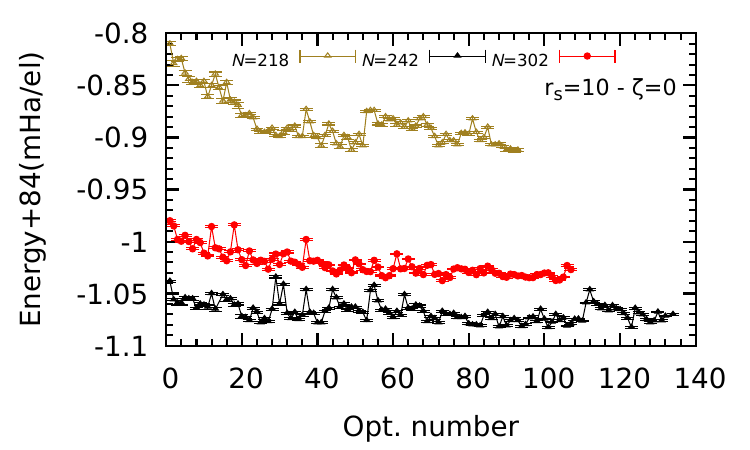} \\
    \includegraphics[scale=0.69]{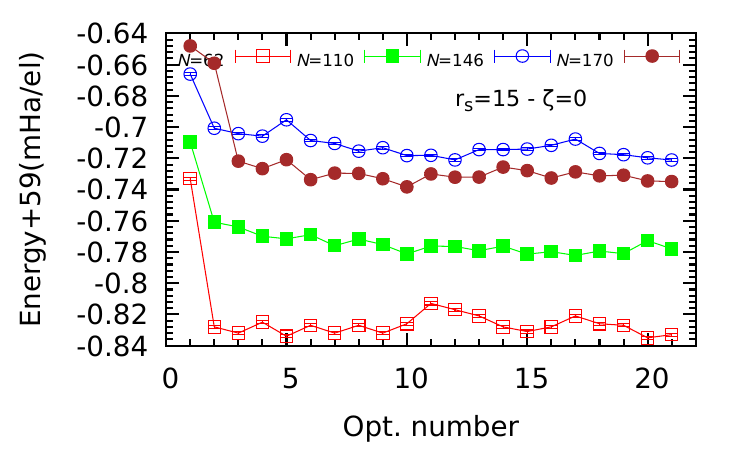} &
    \includegraphics[scale=0.69]{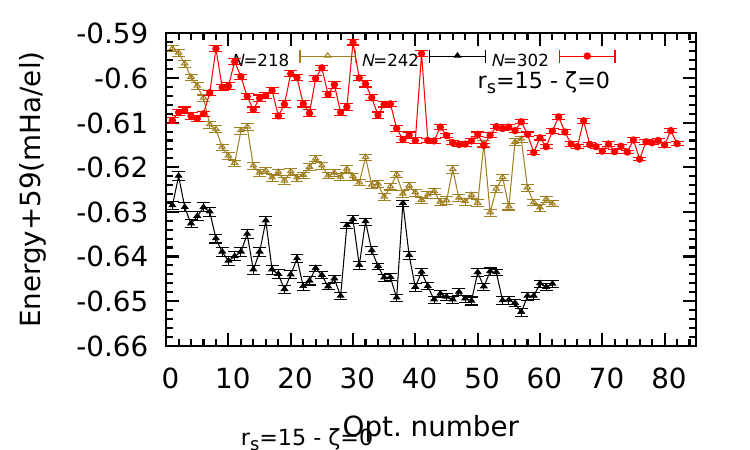} \\
    \includegraphics[scale=0.69]{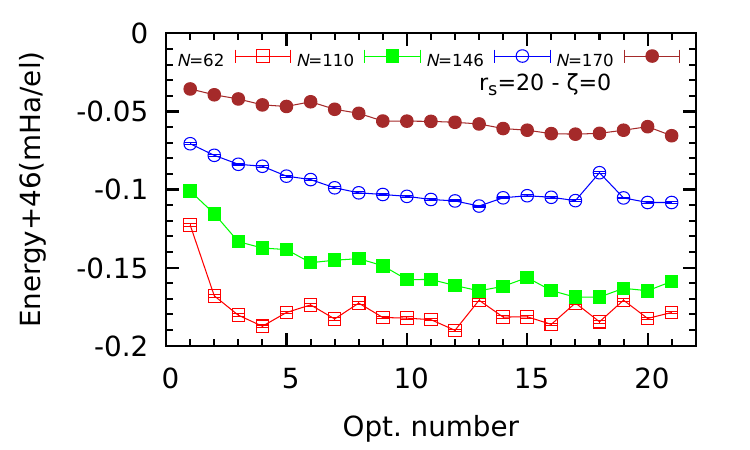} &
    \includegraphics[scale=0.69]{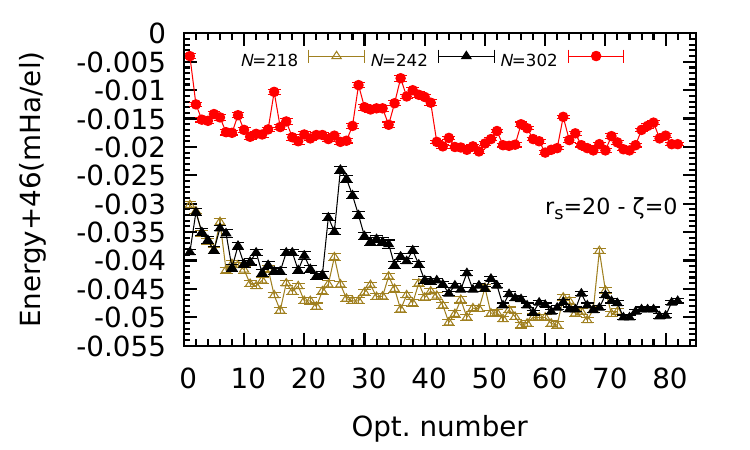} \\
    \includegraphics[scale=0.69]{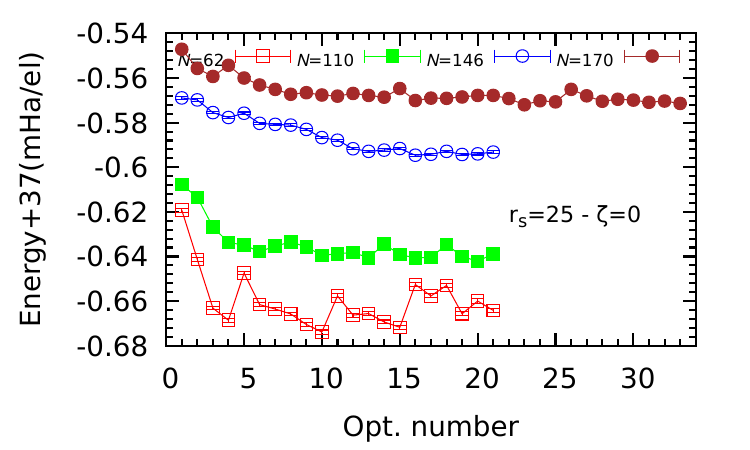} &
    \includegraphics[scale=0.69]{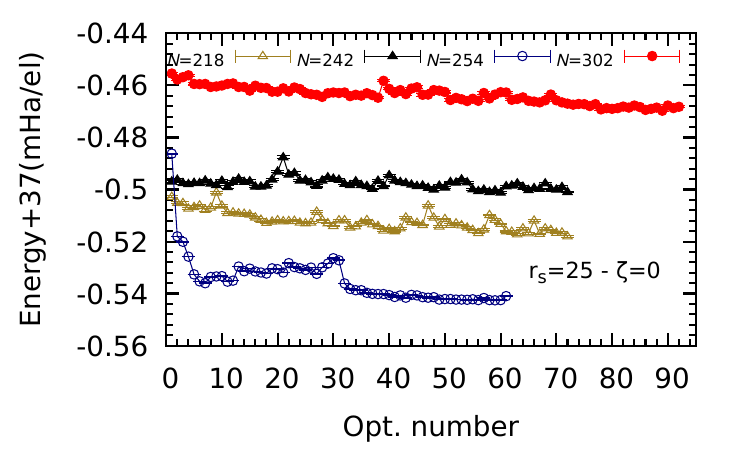} \\
            \end{tabular}
    \caption{Slater-Jastrow-backflow variational quantum Monte Carlo
      total energy against energy minimization cycle for different
      system sizes and density parameters for the paramagnetic
      two-dimensional uniform electron liquid at zero twist. The
      initial wave function was optimized by either unreweighted
      variance minimization or minimization of the mean absolute
      deviation from the median local energy.}
    \label{fig:eminrs10_25}
\end{figure}

\begin{figure}[htbp!]
        \centering
        \begin{tabular}{cc}
    \includegraphics[scale=0.69]{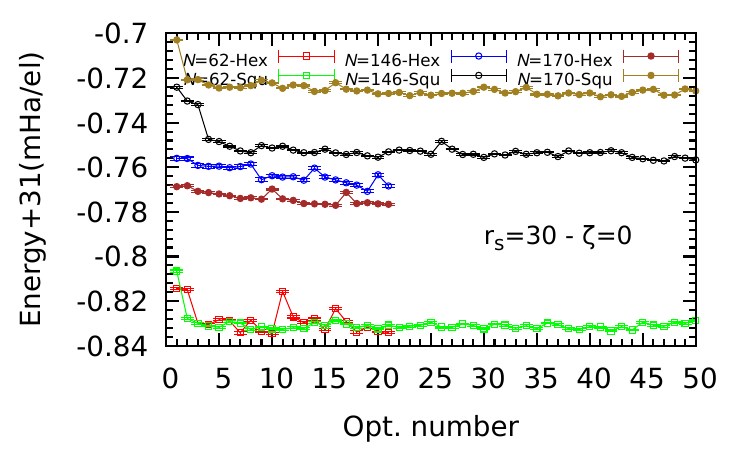} &
    \includegraphics[scale=0.69]{emin_rs30_Nabove200.pdf} \\
    \includegraphics[scale=0.69]{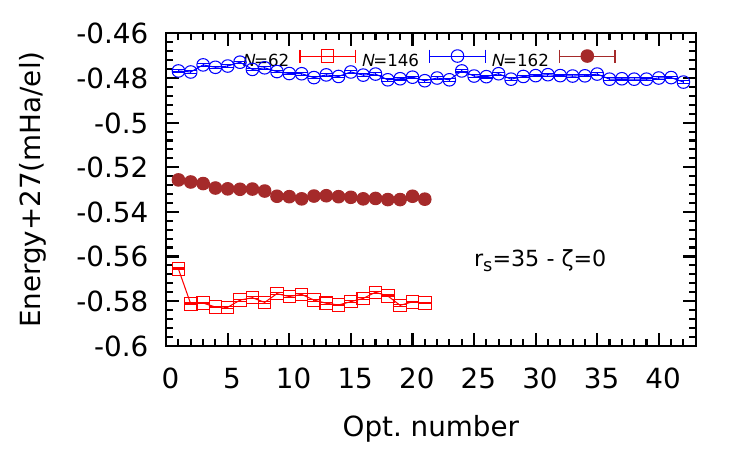} &
    \includegraphics[scale=0.69]{emin_rs35_Nabove200.pdf} \\
            \end{tabular}
    \caption{As Fig.\ \ref{fig:eminrs10_25}, but for lower
      densities. In the case of $r_\text{s}=30$, the energies are
      calculated using hexagonal (Hex) and square (Squ) simulation
      cells.}
    \label{fig:eminrs30_35}
\end{figure}

\begin{table}[htbp!]
    \centering
    \caption{Twist-averaged diffusion Monte Carlo energy
      $E_\text{DMC}$ for $N$-electron two-dimensional uniform electron
      liquids with spin polarization $\zeta=0$ and density parameter
      $r_\text{s}$. $N=\infty$ represents the energy in the limit of
      infinite system size. Slater-Jastrow-backflow wave functions
      were used. \label{tab:paraenergies}}
    \begin{tabular}{lcc|lcc}
    \hline\hline
    $r_\text{s}$  & $N$ & $E_\text{DMC}$ (Ha/el.) & $r_\text{s}$  & $N$
    & $E_\text{DMC}$ (Ha/el.)  \\
    \hline
    1 & 146 & $-0.211457(5)$      & 20 & 62  & $-0.046445(1)$ \\
    1 & 218 & $-0.210922(5)$      & 20 & 110 & $-0.0464184(8)$ \\
    1 & 302 & $-0.210708(6)$      & 20 & 146 & $-0.0464077(8)$ \\
    1 & $\infty$ & $-0.21017(6)$  & 20 & 218 & $-0.046401(1)$  \\
    \ldots&  \ldots& \ldots     & 20 & 242 & $-0.046399(1)$ \\
    2 & 146& $-0.25901(6)$        & 20 & $\infty$ & $-0.046388(1)$ \\
    2 & 218& $-0.258831(6)$       & \ldots & \ldots &\ldots \\
    2 & 302& $-0.258665(5)$       & 25  & 62  & $-0.0378521(7)$  \\
    2 & $\infty$ & $-0.25846(7)$  & 25  & 110 & $-0.0378329(4)$  \\   
    \ldots&  \ldots& \ldots     & 25  & 146 & $-0.0378238(5)$ \\
    5 & 146& $-0.150113(2)$       & 25  & 254 & $-0.0378136(5)$  \\
    5 & 218& $-0.150069(2)$       & 25 & $\infty$ & $-0.037807(3)$\\
    5 & 302& $-0.150024(4)$       &\ldots & \ldots &\ldots\\
    5 & $\infty$ & $-0.14998(2)$  & 30 & 62  & $-0.0319681(9)$ \\  
    \ldots & \ldots& \ldots     & 30 & 170 & $-0.0319466(5)$  \\
    10 &62  & $-0.085709(2)$      & 30 & 254 & $-0.0319435(4)$  \\
    10 &110 & $-0.085643(2)$      & 30 & $\infty$ & $-0.0319383(2)$  \\
    10 &170 & $-0.085612(1)$      & \ldots  &  \ldots  &\ldots \\
    10 &302 & $-0.085582(3)$      & 35 & 62 & $-0.0276932(8)$ \\
    10 & $\infty$ & $-0.085568(5)$& 35 & 122& $-0.0276811(3)$ \\
    \ldots &\ldots &\ldots      & 35 & 254& $-0.0276751(7)$         \\
    15 & 62 & $-0.060214(1)$      & 35 & $\infty$ & $-0.0276718(4)$\\
    15 & 100& $-0.060177(1)$      & \ldots & \ldots  & \ldots\\
    15 & 170& $-0.060160(1)$     & 40  & 62 & $-0.0244424(2)$\\
    15 & 242& $-0.060148(2)$     & 40  & 110& $-0.0244321(2)$ \\
    15 & $\infty$ & $-0.06138(2)$& 40  & 146& $-0.0244295(2)$ \\
    \ldots & \ldots  & \ldots  & 40  & $\infty$& $-0.0244226(3)$ \\
    \hline\hline
    \end{tabular}
   \end{table}
   
\begin{table}[htbp!]
    \centering
    \caption{As Table \ref{tab:paraenergies}, but for a fully
      ferromagnetic two-dimensional uniform electron liquid with spin
      polarization $\zeta=1$. \label{tab:ferroenergies}}
    \begin{tabular}{ccc|ccc}
    \hline\hline
    $r_\text{s}$  & $N$ & $E_\text{DMC}$ (Ha/el.) & $r_\text{s}$  & $N$
    & $E_\text{DMC}$ (Ha/el.)  \\
    \hline
    1 & 55 & $0.124298(6)$         & 20 & 55 & $-0.0462735(5)$ \\
    1 & 91 & $0.125157(5)$         & 20 & 91 & $-0.0462584(4)$ \\
    1 & 169 & $0.125782(2)$        & 20 & 139 & $-0.0462488(2)$ \\
    1 & $\infty$ & $0.12626(4)$    & 20 & 169 & $-0.0462449(2)$ \\
    \ldots & \ldots & \ldots       & 20 & $\infty$& $-0.046236(1)$ \\
    2 & 55 & $-0.195569(3)$        & \ldots & \ldots&\ldots \\
    2 & 91 & $-0.195089(2)$        & 25 & 55  & $-0.0377815(4)$ \\
    2 & 139& $-0.194848(1)$        & 25 & 91  & $-0.0377695(4)$  \\
    2 & 169& $-0.194792(1)$        & 25 & 139 & $-0.0377639(2)$ \\
    2 & $\infty$ & $-0.19453(1)$   & 25 & 169 & $-0.0377603(2)$ \\
    \ldots & \ldots & \ldots       & 25 & $\infty$ & $-0.037754(1)$ \\
    5 & 55 & $-0.143992(2)$        & \ldots & \ldots&\ldots \\
    5 & 91 & $-0.143865(1)$        & 30 & 55 & $-0.0319418(3)$   \\
    5 & 139& $-0.1438037(8)$       & 30 & 91 & $-0.0319340(2)$  \\
    5 & 169& $-0.1437819(5)$       & 30 & 139& $-0.0319296(2)$ \\
    5 & $\infty$ & $-0.143714(2)$  & 30 & 169& $-0.0319243(1)$ \\
    \ldots & \ldots & \ldots       & 30 & $\infty$ & $-0.031919(2)$ \\
    10 & 55 & $-0.084688(1)$       & \ldots & \ldots & \ldots \\
    10 & 91 & $-0.0846419(7)$      & 35 & 55 & $-0.0276857(3)$ \\
    10 & 139 & $-0.0846251(4)$     & 35 & 91 & $-0.0276788(2)$ \\
    10 & 169 & $-0.0846088(4)$     & 35 & 139& $-0.0276739(1)$ \\
    10 & $\infty$ & $-0.084585(6)$ & 35 & 169& $-0.0276719(1)$ \\
    \ldots & \ldots & \ldots       & 35 & $\infty$& $-0.027668(7)$ \\ 
    15 & 55  & $-0.0597888(9)$     & \ldots & \ldots&\ldots \\
    15 & 91  & $-0.0597629(5)$     & 40 &  91 & $-0.0244397(2)$ \\
    15 & 139 & $-0.0597489(4)$     & 40 & 139 & $-0.0244353(1)$ \\
    15 & 169 & $-0.0597451(3)$     & 40 & 169 & $-0.02443388(9)$ \\
    15 & $\infty$ & $-0.059731(1)$ & 40 & $\infty$ & $-0.0244289(1)$ \\
        \hline\hline
    \end{tabular}
\end{table}

\begin{figure}[htbp!]
        \centering
        \begin{tabular}{cc}
    \includegraphics[scale=0.5]{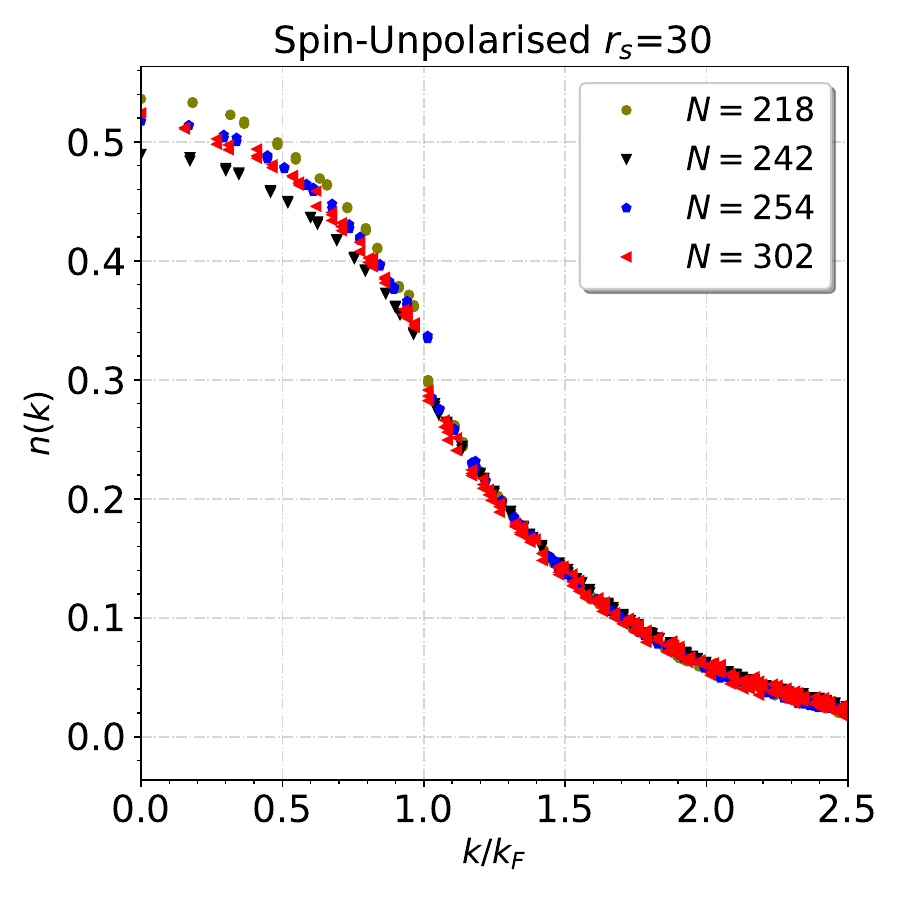} &
    \includegraphics[scale=0.5]{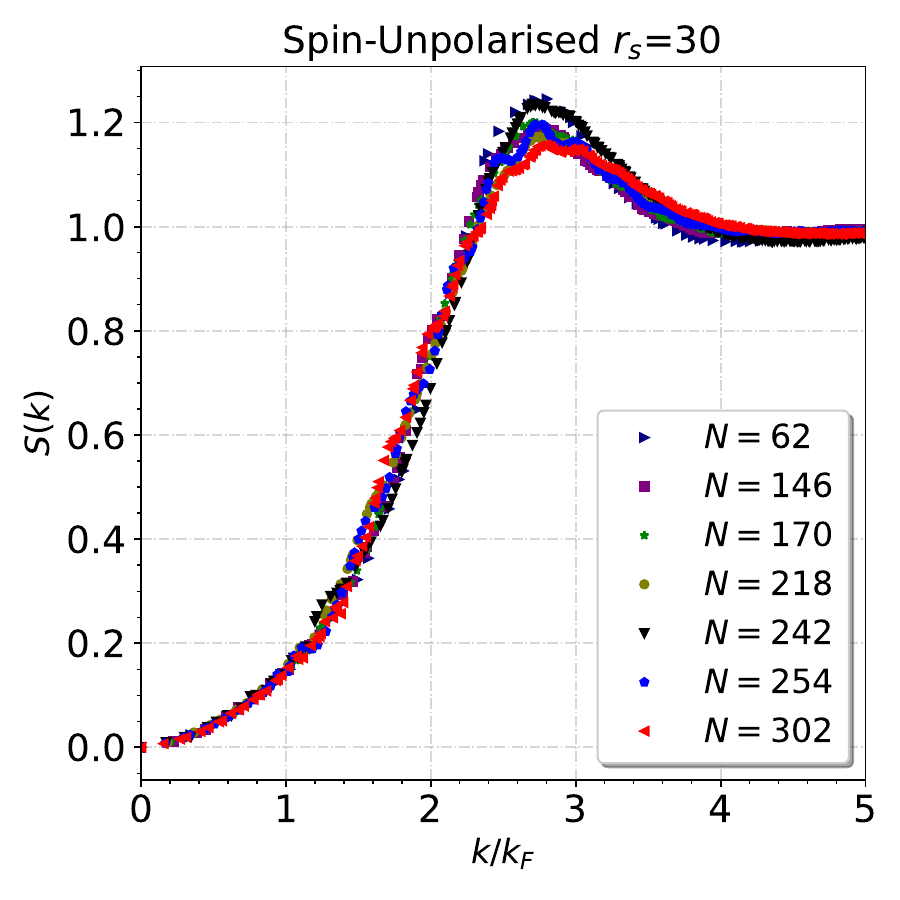} \\
            \end{tabular}
    \caption{Momentum density $n(k)$ and static structure factor
      $S(k)$ of paramagnetic two-dimensional uniform electron liquids using SJB WF
      with density parameter $r_\text{s}=30$ at different system sizes
      $N$ obtained by variational quantum Monte Carlo at zero twist.
      $k_\text{F}$ is the Fermi wavevector.}
    \label{fig:MDandSF}
\end{figure}

\begin{table}[htbp!]
\centering
\caption{Twist-averaged SJB-WF diffusion quantum Monte Carlo energies
  $E_\text{DMC}^\text{Hex}$ and $E_\text{DMC}^\text{Squ}$ of
  two-dimensional uniform electron liquids at $r_\text{s}=30$ obtained
  using hexagonal and square simulation cells.  $\delta E =
  E_\text{DMC}^\text{Hex} - E_\text{DMC}^\text{Squ}$.}
\begin{tabular}{lccc}
 \hline\hline
 $N$ & $E_\text{DMC}^\text{Hex}$ (Ha/el)& $E_\text{DMC}^\text{Squ}$ (Ha/el)
& $\delta E$ (mHa/el.) \\
 \hline
 62   & $-0.0319681(9)$& $-0.0319468(7)$& $-0.021(1)$\\
 146  & $-0.0319262(7)$& $-0.0319272(9)$&  $0.001(1)$\\
 170  & $-0.0319466(5)$& $-0.0319219(8)$& $-0.0247(9)$\\
 218  & $-0.031916(1)$ & $-0.031911(1)$ & $-0.004(2)$\\
 242  & $-0.031904(3)$ & $-0.031901(1)$ & $-0.003(4)$\\
 254  & $-0.0319435(4)$& $-0.0318921(8)$& $-0.051(1)$\\
 302  & $-0.031882(1)$ & $-0.031869(2)$ & $-0.013(2)$\\
 \hline\hline
 \end{tabular}

\end{table}